\documentclass[%
notitlepage,
 amsmath,amssymb,
 aps,
]{revtex4-1}

\usepackage{graphicx}
\usepackage{dcolumn}
\usepackage{bm}
\usepackage{braket}
\usepackage{color}

\begin{document}


\title{Few-photon scattering and emission from open quantum systems}

\author{Rahul Trivedi}
 \email{rtrivedi@stanford.edu}
\author{Kevin Fischer}
\author{Shanshan Xu}
\author{Shanhui Fan}
\author{Jelena Vuckovic}%

\affiliation{%
E. L. Ginzton Laboratory, Stanford University, Stanford, CA 94305, USA }%


\begin{abstract}
We show how to use the input-output formalism compute the propagator for an open quantum system, i.e.~quantum networks with a low dimensional quantum system coupled to one or more loss channels. The total propagator is expressed entirely in terms of the Green's functions of the low dimensional quantum system, and it is shown that these Green's functions can be computed entirely from the evolution of the low-dimensional system with an effective non-hermitian Hamiltonian. Our formalism generalizes the previous works that have focused on time independent Hamiltonians to systems with time dependent Hamiltonians, making it a suitable computational tool for the analysis of a number of experimentally interesting quantum systems. We illustrate our formalism by applying it to analyze photon emission and scattering from driven and undriven two-level system and three-level lambda system.
\end{abstract}

\pacs{Valid PACS appear here}
\maketitle
 

\section{\label{sec_intro}Introduction}
\noindent {Single photon sources \cite{lodahl2015interfacing,ding2016demand,michler,senellart2017high,zhang2018strongly} and single and two-photon optical gates \cite{kok2007linear,o2009photonic,roy2017colloquium,reiserer2015cavity,duan2010colloquium,sangouard2011quantum,nemoto2014photonic} are the basic building blocks of optical quantum information processing and quantum communication systems. Implementing any of these building blocks involves interfacing a low-dimensional quantum system (e.g.~a two level system such as quantum dots, color centers, superconducting qubits, atomic ensembles) with the high-dimensional optical field (which often propagates in optical structures such as waveguides) -- the resulting system is called an open-quantum system \cite{gardiner2004quantum,gardiner2015quantum,wiseman2009quantum} and such systems have been a topic of theoretical interest since the inception of quantum optics. Analyzing open-quantum systems has always been a challenging task due to the huge and continuous Hilbert Space of the optical fields, and the non-linearity induced by the low-dimensional quantum system. 

Traditional computational methods for analyzing open quantum systems fall under two broad classes -- master equation based methods and the scattering matrix based methods. The master-equation based methods \cite{breuer2002theory,carmichael2009open,isar1994open} exactly compute the dynamics of the low-dimensional system while tracing out the Hilbert space of the optical fields. Within the master-equation framework, it is only possible to computationally extract the correlators in the optical fields. To extract the full state of the optical field, correlators of arbitrary orders are required and computing them becomes exponentially more complex with the order of the correlator. The scattering matrix based methods \cite{xu2015input,xu2017input,fischer2017scattering,pletyukhov2015quantum,pan2016exact,caneva2015quantum,shi2015multiphoton,kiukas2015equivalence,fan2010input, shi2011two, shen2005coherent, xu2013analytic, shen2007strongly} attempt to resolve this problem by treating the low-dimensional system as a scatterer for the optical fields and attempting to relate the incoming and outgoing optical fields. However, most of the scattering matrix methods are restricted to time-independent Hamiltonians -- in particular, they are unable to analyze coherently driven systems which are extremely important from an experimental standpoint. Moreover, the scattering matrix methods often only relate the input state at distant past to the output state at distant future, and don't have the capability to model the dynamics of the system during the interaction of the optical field with the low-dimensional system.

In this paper, we present a full calculation of the few-photon elements of the propagator for an arbitrary open quantum system coupled to a optical field. The central result of this paper is a relation between the propagator and time-ordered expectations of system operators over states where the bath is in the vacuum state (labeled as the `Green's functions'). By resorting to a discrete approximation of the bath Hilbert space, we show that Green's functions can be evaluated by computing the dynamics of the low-dimensional system under an effective non-Hermitian Hamiltonian. It can be noted that our formalism is valid for both time-independent and time-dependent Hamiltonians, allowing it to efficiently model not only few-photon transport through the low-dimensional system, but also photon-emission and scattering from coherently driven systems. Our formalism provides a set of computational tools that are complementary to the master-equation framework, wherein the focus is on exactly computing the dynamics of the few-photon states emitted from the low-dimensional system, as opposed to capturing the exact dynamics of the low-dimensional system. Our formalism will thus be of relevance to design and analysis of quantum information processing systems in which the `information' is encoded in the state of the bath, with the low-dimensional system implementing either a source for the bath state or a unitary operation on the bath state.

Finally, we show how to extract the scattering matrices of the low-dimensional system from the full propagator. In particular, we show that for the case of a time-independent low-dimensional system Hamiltonian, our method reduces to scattering matrices derived in past works \cite{xu2015input,xu2017input}. Our formalism also allows us to define a scattering matrix for systems with time-dependent Hamiltonians, as long as they are asymptotically time independent. Such a quantity might be of interest while analyzing photon scattering from coherently driven low-dimensional systems.

This paper is organized as follows -- Section \ref{sec:out_state} describes the propagator computation starting from the input-output equations, Section \ref{sec:scat_matrix} shows how to extract the scattering matrix for the system from the propagator and Section \ref{sec_examples} uses the formalism developed in sections II and III to analyze scattering and emission from a two-level system and a lambda system.}

\section{\label{sec:out_state}Quantum optical systems coupled to waveguides}

\begin{figure}
\includegraphics[scale = 0.9]{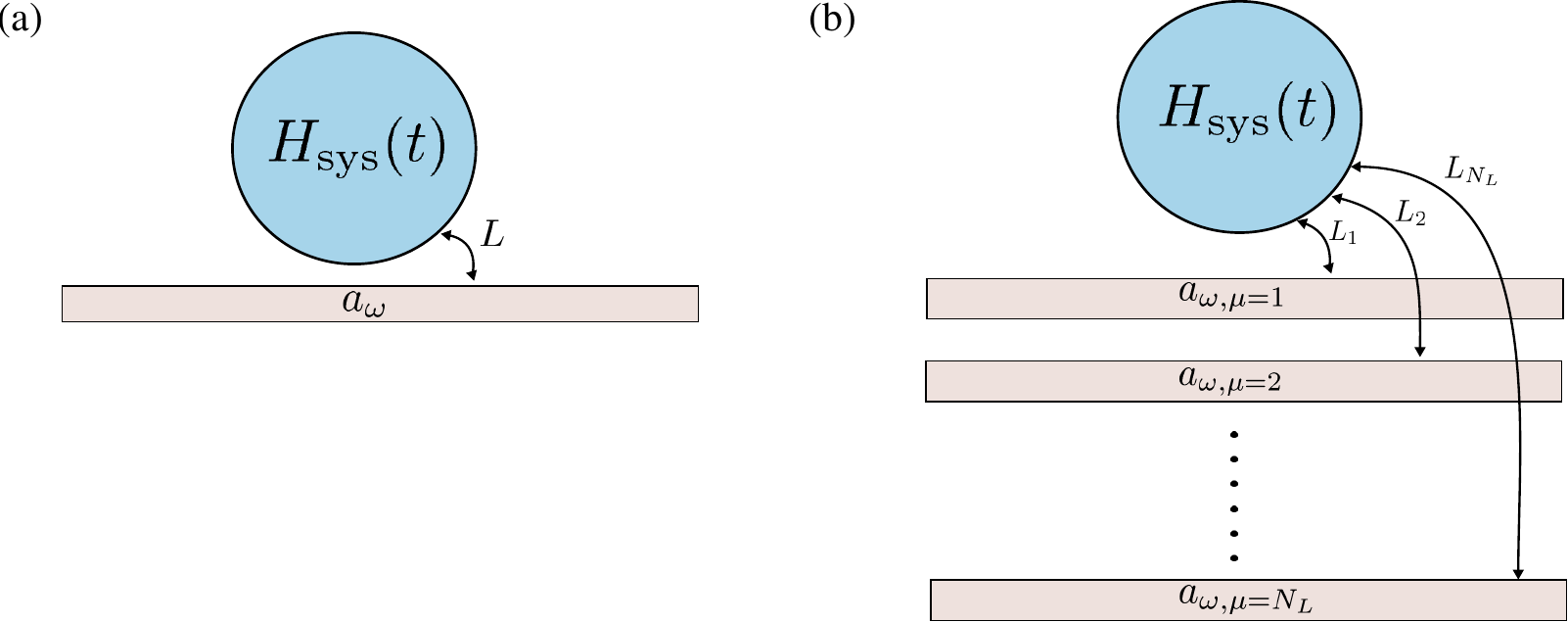}
\caption{Schematic of the quantum systems being considered, with a low-dimensional system coupled to (a) a single waveguide (b) multiple waveguides. }
\label{fig:schematic}
\end{figure}

\noindent {The general quantum networks being considered in this paper include a low dimensional quantum-optical system coupled to one or more waveguides (Fig.~\ref{fig:schematic}). In our calculations, we will be concerned with not only the state of the quantum-optical system, but also with the state of the waveguides. Section \ref{sec:prob-setup} introduces the Hamiltonians studied in this paper and the associated input-output equations, section \ref{sec:inp-out} shows how to use the input-output formalism to relate the propagator matrix elements to Green's functions of the low dimensional system, \ref{sec:mult-wg} extends the result to multiple input and output waveguides and \ref{sec:green-func} shows how to efficiently compute the Green's functions required for computing the propagator matrix elements.}
\subsection{\label{sec:prob-setup}Hamiltonian and input-output formalism}
\noindent As a starting point, we consider only one waveguide [Fig.~\ref{fig:schematic}(a)] --- the generalization to multiple waveguides is a straightforward extension. The total Hamiltonian we are interested in can be expressed as the local system Hamiltonian $H_\text{sys}$, the waveguide Hamiltonian $H_\text{wg}$ and an interaction Hamiltonian between the local system and waveguide $H_\text{int}$:
\begin{equation}
H = H_\text{sys}(t)+H_\text{wg}+H_\text{int}.
\end{equation}
The Hilbert space of the low dimensional system can often be captured by a finite or countably infinite basis, which we denote by $\{\ket{\sigma_1}, \ket{\sigma_2} \dots \ket{\sigma_S}  \}$. It is typically straightforward and computationally tractable to analyze the dynamics of the low-dimensional system in isolation. 

The dynamics of the waveguide can be modeled by the following Hamiltonian:
\begin{equation}
H_\text{wg} = \int \omega a_\omega^\dagger a_\omega \textrm{d}\omega
\end{equation}
where $a_\omega$ is the annihilation operator for an excitation of the waveguide mode at frequency $\omega$. They satisfy the standard commutation relation $[a_\omega, a_{\omega'}] = 0$ and $[a_\omega, a_{\omega'}^\dagger] = \delta(\omega-\omega')$. A suitable basis for the waveguide's Hilbert space is the Fock state basis:
\begin{equation}
\ket{\omega_1, \omega_2 \dots \omega_N} = \prod_{n = 1}^N a_{\omega_n}^\dagger \ket{\text{vac}} \quad\forall N\in\mathbb{N}.
\end{equation}
It is also convenient to define a spatial annihilation operator for the waveguide mode:
\begin{equation}
a_x = \int a_\omega \exp \bigg(\textrm{i} \frac{\omega x}{v_g}\bigg) \frac{\textrm{d}\omega}{\sqrt{2\pi v_g}}
\end{equation}
where $v_g$ is the group velocity of the waveguide mode under consideration, which we will take as unity ($v_g=1$) for the rest of this paper (note that this is equivalent to rescaling the position $x$ to have units of time). It follows from the commutators for the operators $a_\omega$ that $[a_x,a_{x'}] = 0$ and $[a_x,a^\dagger_{x'}] = \delta(x-x')$. These operators can be used to construct another basis for the waveguide's Hilbert space, which we refer to as the spatial Fock states:
\begin{equation}
\ket{x_1, x_2 \dots x_N} = \prod_{n=1}^N a_{x_n}^\dagger \ket{\text{vac}}.
\end{equation}

The system-waveguide interaction, in the rotating wave approximation, can be described by the following Hamiltonian:
\begin{equation}
H_\text{int} = \int \big(a_\omega L^\dagger +L a_\omega^\dagger \big)\,  \frac{\textrm{d}\omega}{\sqrt{2\pi}}
\end{equation}
where $L$ is the operator through which the low-dimensional system couples to the waveguide. In writing this Hamiltonian, we make the standard Markov approximation of assuming that the low-dimensional system couples equally to waveguide modes at all frequency --- this is equivalent to assuming that the physical process inducing an interaction between the waveguide and the low-dimensional system has a bandwidth much larger than any excitation that will be used to probe the low-dimensional system.

In the Heisenberg picture, this system can be modeled via well known input-output equations \cite{gardiner1985input}:
\begin{widetext}
\begin{subequations} \label{in_out_eq}
\begin{align}
&a_\omega(t) = a_\omega(\tau_-)\exp(-\textrm{i}\omega (t-\tau_-)) 
-\textrm{i} \int_{\tau_-}^t L(t')\exp(-\textrm{i}\omega(t-t')) \,\frac{\textrm{d}t'}{\sqrt{2\pi}} \label{bath_op}\\
&\dot{L}(t) = -\textrm{i}[L(t),H_\text{sys}]- \textrm{i} [L(t),L^\dagger(t)]\bigg(a_\text{in}(t)-\frac{\textrm{i}}{2}L(t) \bigg)
\end{align}
\end{subequations}
\end{widetext}
where $a_\text{in}(t) = \int a_\omega(\tau_-)\exp(-\textrm{i}\omega (t-\tau_-)) \,\textrm{d}\omega/\sqrt{2\pi}$ is the input operator which, in the Heisenberg picture, captures the state of the system at time $t = \tau_-$. After the evolution of the system to a time instant $t = \tau_+>\tau_-$ in the future, the state of the system is described by the output operator $a_\text{out}(t) = \int a_\omega(\tau_+)\exp(-\textrm{i}\omega(t-\tau_+))\,\textrm{d}\omega/\sqrt{2\pi}$. Using Eq.~\ref{bath_op}, we can relate the input operator to the output operator via: 
\begin{align}
a_\text{out}(t) = a_\text{in}(t) - \textrm{i} L(t) \mathcal{I}(\tau_-<t<\tau_+)\label{eq:inout}
\end{align}
where $\mathcal{I}(\cdot)$ is the indicator function, which is 1 if its argument is true or else is 0. Intuitively, this expresses the field in the waveguide after the input has interacted with the system as a sum of the input field and field emitted by the low-dimensional system.

A useful set of commutators between the input operator, output operator and the low-dimensional system's Heisenberg operators can be derived using the quantum-causality conditions \cite{gardiner1985input}:
\begin{align}
[a_\text{out}(t), s(t')] = 0 \text{ if } t<t' \ \text{and} \ [a_\text{in}(t), s(t')] = 0 \text{ if } t>t'
\end{align}
where $s$ is an operator in the Hilbert space of the low-dimensional system. From these causality conditions, it immediately follows that \cite{xu2015input}:
\begin{subequations}
\begin{align}
\mathcal{T}\bigg[\prod_{l=1}^L o(t_l) \prod_{m=1}^M s_m(t_m') \bigg] = \mathcal{T}\bigg[\prod_{l=1}^L o(t_l)\bigg] \mathcal{T}\bigg[\prod_{m=1}^M s_m(t_m') \bigg]\label{eq:time_order_output}\\
\mathcal{T}\bigg[\prod_{l=1}^L i(t_l) \prod_{m=1}^M s_m(t_m') \bigg] =  \mathcal{T}\bigg[\prod_{m=1}^M s_m(t_m') \bigg]\mathcal{T}\bigg[\prod_{l=1}^L i(t_l)\bigg]\label{eq:time_order_input}
\end{align}
\end{subequations}
where $o(t)$ is an output operator evaluated at time $t$ (i.e.~$a_\text{out}(t)$, $a_\text{out}^\dagger(t)$ or their combination), $i(t)$ is an input operator evaluated at time $t$ (i.e.~$a_\text{in}(t)$, $a_\text{in}^\dagger(t)$ or their combination), $s_m(t)$ is a low-dimensional system operator evaluated at time $t$ in the Heisenberg picture and $\mathcal{T}$ is the chronological time-ordering operator.

\subsection{\label{sec:inp-out}Calculating the propagator matrix elements\label{sec:calcprop}}
\noindent The dynamics of a quantum system with a Hamiltonian $H(t)$ is completely characterized by its propagator -- in particular, if the propagator is known, then the time evolution of the quantum state of the system can be completely derived from the initial state of the system. In this section, we will focus on computing the interaction picture propagator $U_I(\tau_{+},\tau_{-})$  defined by:
\begin{align}
U_\textrm{I}(\tau_{+},\tau_{-}) = \exp(\textrm{i}H_\text{wg}\tau_+)U(\tau_+,\tau_-)\exp(-\textrm{i}H_\text{wg}\tau_-) 
\end{align}
where $U(\tau_+,\tau_-)$ is the Schr\"{o}dinger picture propagator for the system:
\begin{equation}
U(\tau_{+},\tau_{-}) = \mathcal{T}\text{exp}\left[-\textrm{i}\int_{\tau_-}^{\tau_+} H(t) \textrm{d}t\right]\label{eq:U}
\end{equation}
and $\mathcal{T}$ is the chronological operator that time-orders the infinitesimal products of Eq. \ref{eq:U}.
In particular, we will compute the matrix elements of the interaction picture propagator in the form:
\begin{align}
U^{\sigma', \sigma}_{\tau_+,\tau_-}(x_1', x_2' \dots x_M'; x_1, x_2 \dots x_N) \equiv \bra{x_1', x_2' \dots x_M';\sigma'} U_\textrm{I}(\tau_+,\tau_-) \ket{x_1, x_2 \dots x_N; \sigma}
\end{align}
where $\ket{x_1, x_2 \dots x_P; \sigma} = \ket{x_1, x_2 \dots x_P} \otimes \ket{\sigma}$ with $x_i \in \mathbb{R}$ and $\sigma \in \{\sigma_1, \sigma_2 \dots\}$. Since the spatial Fock state basis for the waveguide is complete, and the basis $\{\ket{\sigma_1}, \ket{\sigma_2}\dots \}$ is a complete basis for the system state by construction, these matrix elements are sufficient to characterize the entire propagator.

Writing out $U^{\sigma', \sigma}_{\tau_+,\tau_-}(x_1', x_2' \dots x_M'; x_1, x_2 \dots x_N)$ in terms of the spatial annihilation operators:
\begin{align}
U^{\sigma', \sigma}_{\tau_+,\tau_-}(x_1' \dots x_M'; x_1 \dots x_N) &= \bra{\text{vac}; \sigma'}\bigg[\prod_{i=1}^M a_{x_i'}  \bigg] U_\textrm{I}(\tau_+,\tau_-) \bigg[ \prod_{j=1}^N a_{x_j}^\dagger\bigg ]\ket{\text{vac}; \sigma} \nonumber \\
&=  \bra{\text{vac}; \sigma'}\bigg[\prod_{i=1}^M a_{x_i'}  \bigg] U_\textrm{I}(\tau_+,0)U(0,\tau_-) \bigg[ \prod_{j=1}^N a_{x_j}^\dagger\bigg ]\ket{\text{vac}; \sigma}
\label{eq:U(x',x)}.
\end{align}
Noting that
\begin{equation}
\exp(-\textrm{i}H_\text{wg}\tau) a_x \exp(\textrm{i}H_\text{wg}\tau) = \int a_\omega \exp(\textrm{i}\omega(x+\tau)) \frac{\textrm{d}\omega}{\sqrt{2\pi}}
\end{equation}
and since $U(0,\tau) a_\omega U(\tau,0) = a_\omega(\tau)$,
\begin{equation}\label{eq:out_spatial}
U_\textrm{I}(0,\tau_+) a_x U_\textrm{I}(\tau_+,0) = a_\text{out}(-x) \; \text{ and } \; U_\textrm{I}(0,\tau_-) a_x U_\textrm{I}(\tau_-,0) = a_\text{in}(-x).
\end{equation}
Therefore
\begin{subequations}
\begin{align}
&\bigg[\prod_{i=1}^M a_{x_i'}\bigg] U_\textrm{I}(\tau_+,0) = U_\textrm{I}(\tau_+,0) \bigg[\prod_{i=1}^M a_\text{out}(-x_i')\bigg]  \\
&U_\textrm{I}(0,\tau_-)\bigg[\prod_{j=1}^M a_{x_j}^\dagger\bigg] = \bigg[\prod_{j=1}^M a_\text{in}^\dagger(-x_j)\bigg] U_\textrm{I}(0,\tau_-).
\end{align}
\end{subequations}
Eq.~\ref{eq:U(x',x)} can thus be expressed as
\begin{align}\label{eq:prop_mod}
&U_{\tau_+,\tau_-}^{\sigma',\sigma}(x_1'\dots x_M'; x_1 \dots x_N) = \bra{\text{vac}; \sigma'}U_\textrm{I}(\tau_+,0)\prod_{i=1}^M a_\text{out}(-x_i') \prod_{j=1}^N a_\text{in}^\dagger(-x_j) U_\textrm{I}(0,\tau_-) \ket{\text{vac}; \sigma}.
\end{align}
Since all the input or output operators commute with each other, we can introduce a time ordering operator as shown below:
\begin{align}\label{eq:simp_1}
&\prod_{j=1}^N a_\text{in}^\dagger(-x_j) =\mathcal{T}\bigg[\prod_{j=1}^N a_\text{in}^\dagger(-x_j) \bigg]\nonumber \\
&=\mathcal{T} \bigg[\prod_{j=1}^N \big(a_\text{out}^\dagger(-x_j) -\textrm{i}L^
\dagger (-x_j)\mathcal{I}(-\tau_+<x_j<-\tau_-)\big) \bigg] \nonumber\\
&=\sum_{k=0}^N (-\textrm{i})^{N-k} \sum_{B_k^N}\mathcal{T}\bigg[\prod_{i=1}^k a^\dagger_\text{out}(-x_{B_k^N(i)}) \prod_{j=1}^{N-k} L^\dagger(-x_{\bar{B}_k^N(j)})\bigg] \bigg[\prod_{j=1}^{N-k} \mathcal{I}(-\tau_+<x_{\bar{B}_k^N(j)}<-\tau_-) \bigg] \nonumber \\
&=\sum_{k=0}^N (-\textrm{i})^{N-k} \sum_{B_k^N}\bigg[\prod_{i=1}^k a^\dagger_\text{out}(-x_{B_k^N(i)}) \bigg] \mathcal{T}\bigg[\prod_{j=1}^{N-k} L^\dagger(-x_{\bar{B}_k^N(j)})\bigg] \bigg[\prod_{j=1}^{N-k}\mathcal{I}(-\tau_+<x_{\bar{B}_k^N(j)}<-\tau_-) \bigg] 
\end{align}
where $B_k^N$ is a $k-$element unordered subset of $\{1,2 \dots N\}$ and $\bar{B}_k^N$ is its complement. In the last step, we have used Eq.~\ref{eq:time_order_output} and the fact that output operators evaluated at different time instances commute. From Eq.~\ref{eq:out_spatial}, it follows that $a_\text{out}(-x)U_\text{I}(0,\tau_+)\ket{\text{vac}; \sigma'} = U_\text{I}(0,\tau_+)a_x\ket{\text{vac};\sigma} = 0$. Together with the commutator $[a_\text{out}(t), a_\text{out}^\dagger(t')] = \delta(t-t')$, this can be used to prove that:
\begin{align}\label{eq:simp_2}
&\prod_{i=1}^k a_\text{out}(-x_{B_k^N(i)}) \prod_{j=1}^M a_\text{out}^\dagger(-x_j') U_\text{I}(0,\tau_+) |\text{vac};\sigma'\rangle \nonumber \\
&= \mathcal{I}(M\geq k)\sum_{B_k^M}\bigg[\sum_{P_k}\prod_{i=1}^k \delta(x_{P_k B_k^M(i)}'-x_{B_k^N(i)})\bigg]\prod_{j=1}^{M-k} a_\text{out}^\dagger(-x'_{\bar{B}_k^M(j)}) U_\text{I}(0,\tau_+) \ket{\text{vac};\sigma'}
\end{align}
where $P_k$ denotes a permutation of a set of $k$ elements, e.g. $P_k B_k^M$ is a permutation of $B_k^M$ which itself is an unordered $k$-element subset of $\{1,2 \dots M\}$. Substituting Eqs.~\ref{eq:simp_1} and \ref{eq:simp_2} into Eq.~\ref{eq:prop_mod}, we obtain:
\begin{align}\label{eq:U_almost}
&U_{\tau_+,\tau_-}^{\sigma',\sigma}(x_1'\dots x_M'; x_1 \dots x_N) \nonumber \\
&=\sum_{k=0}^N (-\textrm{i})^{N-k}\mathcal{I}(M\geq k)\sum_{B_k^N, B_k^M} \bigg[\sum_{P_k} \prod_{i=1}^k \delta(x'_{P_k B_k^M(i)}-x_{B_k^N(i)}) \bigg]\times \nonumber \\ 
& \quad\bra{\text{vac}; \sigma'}U_\textrm{I}(\tau_+,0) \bigg[\prod_{j=1}^{M-k}a_\text{out}(-x'_{\bar{B}_k^M(j)})\bigg]\mathcal{T}\bigg[\prod_{s=1}^{N-k} L^\dagger(-x_{\bar{B}_k^N(s)})\bigg]U_\textrm{I}(0,\tau_-) \ket{\text{vac}; \sigma}\times \nonumber \\ & \quad\prod_{s=1}^{N-k}\mathcal{I}(-\tau_+<x_{\bar{B}_k^N(s)}<-\tau_-) \nonumber \\
&=\sum_{k=0}^N (-\textrm{i})^{N-k}\mathcal{I}(M\geq k)\sum_{B_k^N, B_k^M} \bigg[\sum_{P_k} \prod_{i=1}^k \delta(x'_{P_k B_k^M(i)}-x_{B_k^N(i)}) \bigg]\times \nonumber \\ 
& \quad\bra{\text{vac}; \sigma'}U_\textrm{I}(\tau_+,0) \mathcal{T}\bigg[\prod_{j=1}^{M-k}a_\text{out}(-x'_{\bar{B}_k^M(j)}) \prod_{s=1}^{N-k} L^\dagger(-x_{\bar{B}_k^N(s)})\bigg]U_\textrm{I}(0,\tau_-) \ket{\text{vac}; \sigma}\times  \nonumber \\ &\quad\prod_{s=1}^{N-k}\mathcal{I}(-\tau_+<x_{\bar{B}_k^N(s)}<-\tau_-) 
\end{align}
wherein we have used Eq.~\ref{eq:time_order_input} in the last step to pull the output operators into the time-ordering operator. Using Eq.~\ref{eq:inout}:
\begin{align}\label{eq:simp_3}
&\mathcal{T}\bigg[\prod_{j=1}^{M-k}a_\text{out}(-x'_{\bar{B}_k^M(j)}) \prod_{s=1}^{N-k} L^\dagger(-x_{\bar{B}_k^N(s)}) \bigg] U_\textrm{I}(0,\tau_-) \ket{\text{vac}; \sigma} \nonumber \\ 
&= \mathcal{T}\bigg[\prod_{j=1}^{M-k}(a_\text{in}(-x'_{\bar{B}_k^M(j)})-\textrm{i}L(-x'_{\bar{B}_k^M(j)})\mathcal{I}(-\tau_+ < x_{\bar{B}_k^M(k)}'<\tau_-)) \prod_{s=1}^{N-k} L^\dagger(-x_{\bar{B}_k^N(s)})  \bigg] U_\textrm{I}(0,\tau_-) \ket{\text{vac}; \sigma} \nonumber \\
&=(-\textrm{i})^{M-k}\mathcal{T}\bigg[\prod_{j=1}^{M-k}L(-x'_{\bar{B}_k^M(j)}) \prod_{s=1}^{N-k}L^\dagger(-x_{\bar{B}_k^N(s)}) \bigg]U_\textrm{I}(0,\tau_-)\ket{\text{vac};\sigma}\prod_{s=1}^{M-k}\mathcal{I}(-\tau_+ < x_{\bar{B}_k^M(s)}'< -\tau_-)
\end{align}
wherein we have used Eq.~\ref{eq:time_order_input} and the fact that input operators at different time instances commute to move all the input operators in the time ordered product to the right and Eq.~\ref{eq:out_spatial} to set $a_\text{in}(-x)U_I(0,\tau_-)\ket{\text{vac};\sigma} = U_I(0,\tau_-)a_x\ket{\text{vac}; \sigma} = 0 $. Substituting Eq.~\ref{eq:simp_3} into Eq.~\ref{eq:U_almost}
\begin{align}\label{eq:prop_green_func}
&U_{\tau_+,\tau_-}^{\sigma',\sigma}(x_1'\dots x_M'; x_1 \dots x_N) =\sum_{k=0}^N (-\textrm{i})^{M+N-2k}\mathcal{I}(M\geq k) \times \nonumber \\ &\qquad \sum_{B_k^N, B_k^M} \bigg[\sum_{P_k} \prod_{i=1}^k \delta(x'_{P_k B_k^M(i)}-x_{B_k^N(i)}) \bigg] \bigg[\prod_{i=1}^{M-k} \mathcal{I}(-\tau_+<x'_{\bar{B}^M_k(i)}<-\tau_-)\bigg] \bigg[\prod_{j=1}^{N-k}  \mathcal{I}(-\tau_+<x_{\bar{B}^N_k(j)}<-\tau_-)\bigg] \times \nonumber \\
& \qquad\mathcal{G}^{\sigma',\sigma}_{\tau_+,\tau_-}(-x'_{\bar{B}_k^M(1)}, -x'_{\bar{B}_k^M(2)} \dots -x'_{\bar{B}_k^M(M-k)};-x_{\bar{B}_k^N(1)}, -x_{\bar{B}_k^N(2)} \dots -x_{\bar{B}_k^N(N-k)})
\end{align}
where $\mathcal{G}_{\tau_+,\tau_-}^{\sigma',\sigma}(t_1'\dots t_M'; t_1 \dots t_N)$ is the system Green's function defined by:
\begin{align} \label{eq:green_func}
\mathcal{G}_{\tau_+,\tau_-}^{\sigma',\sigma}(t_1'\dots t_M'; t_1 \dots t_N) = \bra{\text{vac};\sigma'}U_I(\tau_+,0)\mathcal{T}\bigg[\prod_{i=1}^M L(t_i') \prod_{j=1}^N L^\dagger(t_j) \bigg]U_I(0,\tau_-)\ket{\text{vac}; \sigma}.
\end{align}
Note that the Green's function depends entirely on the dynamics of the low-dimensional system under consideration --- we have thus reduced the problem of computing the propagator for the entire system to the problem of computing the dynamics of only the low-dimensional system. 

Finally, after having computed the propagator in the interaction picture, $U_I(\tau_+, \tau_-)$, it is a simple matter to compute the propagator in the Schr\"{o}dinger picture $U(\tau_+, \tau_-)$. To do so, we use the following property of the spatial Fock state:
\begin{align}
\exp(-\textrm{i}H_\text{wg} \tau) \ket{x_1, x_2 \dots x_N} = \ket{x_1+\tau, x_2+\tau \dots x_N+\tau}
\end{align}
which intuitively states that an excitation created at a position $x$ in the waveguide propagates along the positive $x$ direction with velocity equal to the group velocity of the waveguide mode (which in this case is taken to be 1). Thus:
\begin{align}
&\bra{x_1'\dots x_M';  \sigma'}U(\tau_+, \tau_-) \ket{x_1 \dots x_N; \sigma} = U_{\tau_+, \tau_-}^{\sigma',\sigma}(x_1'-\tau_+, x_2-\tau_+ \dots x_M'-\tau_+; x_1-\tau_-, x_2-\tau_- \dots x_N-\tau_-).
\end{align}

\subsection{\label{sec:mult-wg}Extension to multiple waveguides}

\noindent Local systems coupled to multiple waveguide modes (which can either be physically separate waveguides or orthogonal modes of the same waveguide), diagrammatically shown in Fig.~\ref{fig:schematic}(b), can be described by Hamiltonians of the form:
\begin{equation}\label{mult_loss_channels}
H = H_\text{sys}(t) + \sum_{\mu =1}^{N_L}\int \omega a_{\omega,\mu}^\dagger a_{\omega,\mu} \textrm{d}\omega +\sum_{\mu=1}^{N_L}\int \big( a_{\omega, \mu} L_\mu^\dagger+L_\mu a_{\omega, \mu}^\dagger \big) \frac{\textrm{d}\omega}{\sqrt{2\pi}} 
\end{equation}
where $N_L$ is the total number of loss channels, $a_{\omega, \mu}$ is the plane wave annihilation operator for the $\mu^\text{th}$ waveguide which couples to the low dimensional system through the operator $L_\mu$. A complete basis for the waveguide modes can now be constructed using the creation operators for the different waveguides:
\begin{align}
|\{x_{1},\mu_1\}, \{x_{2},\mu_2\} \dots \{x_{N}, \mu_N\}\rangle = \prod_{i=1}^N a_{x_i, \mu_i}^\dagger \ket{\text{vac}}
\end{align}
where $a_{x,\mu} = \int a_{\omega, \mu}\exp(\textrm{i}\omega x) \textrm{d}\omega/\sqrt{2\pi}$ is the spatial annihilation operator for the $\mu^\text{th}$ waveguide mode. These annihilation operator have the commutators $[a_{x,\mu}, a_{x',\mu'}] = 0$ and $[a_{x,\mu}, a_{x',\mu'}^\dagger] = \delta(x-x')\delta_{\mu,\mu'}$. The propagator can thus be completely characterized by matrix elements of the form:
\begin{align}
&U_{\tau_+, \tau_-}^{\sigma', \sigma}(\{x_{1}',\mu_1'\},\{x_{2}',\mu_2'\}\dots \{x_{M}',\mu_M'\}; \{x_{1},\mu_1\},\{x_{2},\mu_2\}\dots \{x_{N},\mu_N\}) \nonumber \\
&= \bra{\text{vac}; \sigma'}a_{x_1',\mu_1'}a_{x_2',\mu_2'}\dots a_{x_M',\mu_M'} U_\textrm{I}(\tau_+, \tau_-)a_{x_1,\mu_1}^\dagger a_{x_2,\mu_2}^\dagger \dots a_{x_M,\mu_M}^\dagger \ket{\text{vac};\sigma}.
\end{align}
Repeating the procedure described in Section \ref{sec:calcprop}, it can easily be shown that:
\begin{align}\label{eq:prop_green_func_multiple}
&U_{\tau_+,\tau_-}^{\sigma',\sigma}(\{x_1',\mu_1'\}\dots \{x_M',\mu_M'\}; \{x_1,\mu_1\} \dots \{x_N,\mu_N\}) =\sum_{k=0}^N (-\textrm{i})^{M+N-2k}\mathcal{I}(M\geq k) \times \\ &\qquad\sum_{B_k^N, B_k^M} \bigg[\sum_{P_k} \prod_{i=1}^k \delta(x'_{P_k B_k^M(i)}-x_{B_k^N(i)})\delta_{\mu_{P_k B_k^M(i)}', \mu_{B_k^N(i)}} \bigg] \times \nonumber \\  
&\qquad\bigg[\prod_{i=1}^{M-k} \mathcal{I}(-\tau_+<x_{\bar{B}^M_k(i)}<-\tau_-)\bigg] \bigg[\prod_{j=1}^{N-k}  \mathcal{I}(-\tau_+<x_{\bar{B}^N_k(j)}<-\tau_-)\bigg] \times \nonumber \\
& \qquad\mathcal{G}^{\sigma',\sigma}_{\tau_+,\tau_-}(\{-x'_{\bar{B}_k^M(1)}, \mu_{\bar{B}_k^M(1)}'\}, \dots \{-x'_{\bar{B}_k^M(M-k)},\mu'_{\bar{B}_k^M(M-k)}\};\{-x_{\bar{B}_k^N(1)},\mu_{\bar{B}_k^N(1)}\} \dots \{-x_{\bar{B}_k^N(N-k)},\mu_{\bar{B}_k^N(N-k)}\})\nonumber
\end{align}
where:
\begin{align} 
&\mathcal{G}_{\tau_+,\tau_-}^{\sigma',\sigma}(\{t_1',\mu_1'\} \dots \{t_M',\mu_M'\}; \{t_1,\mu_1\} \dots \{t_N,\mu_N\}) =\bra{\text{vac};\sigma'}U_\textrm{I}(\tau_+,0)\mathcal{T}\bigg[\prod_{i=1}^M L_{\mu_i'}(t_i') \prod_{j=1}^N L_{\mu_j}^\dagger(t_j) \bigg]U_\textrm{I}(0,\tau_-)\ket{\text{vac}; \sigma}.
\end{align}

\subsection{\label{sec:green-func}Efficient computation of the Green's functions}
\begin{figure}
\includegraphics[scale = 0.8]{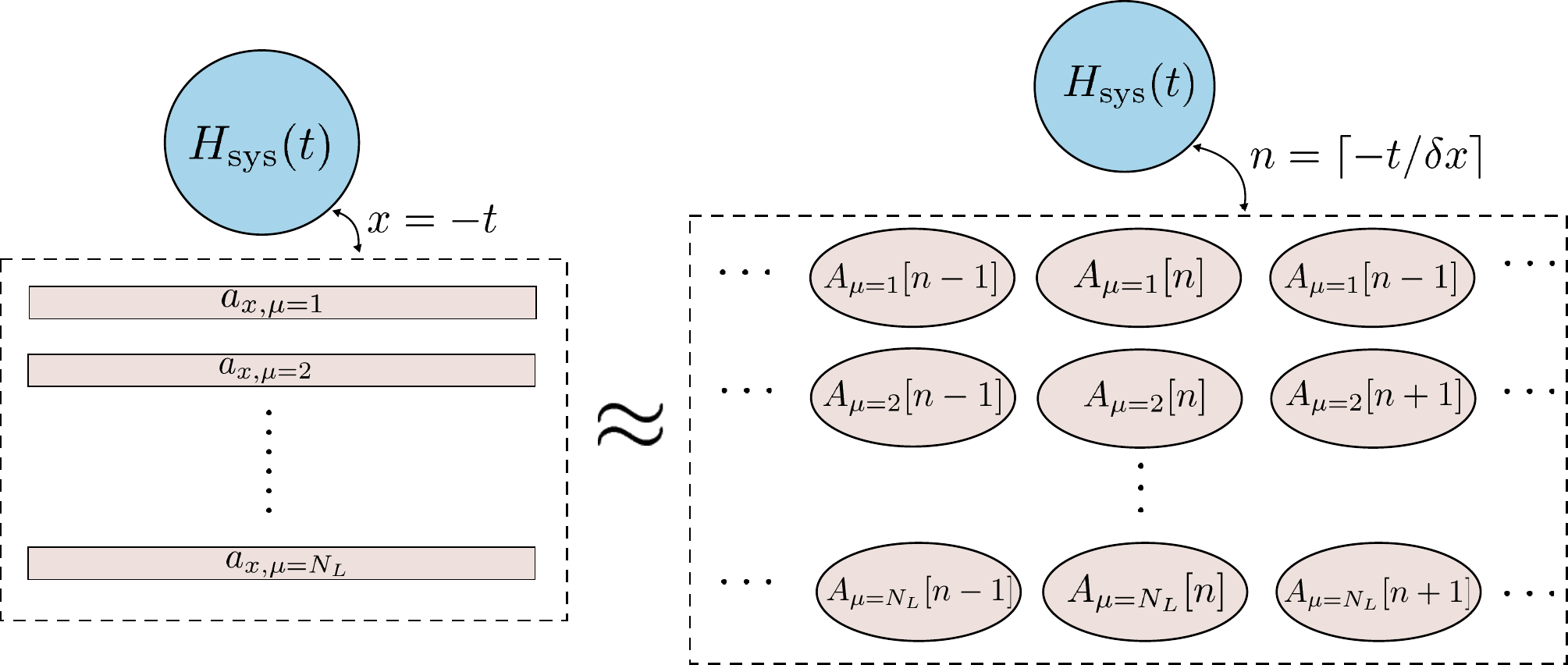}
\caption{Diagrammatic representation of approximating the Hilbert space of the bath modes with a discrete Hilbert space.}
\label{fig:fig_apprx}
\end{figure}
\noindent While Eqs.~\ref{eq:prop_green_func} and \ref{eq:prop_green_func_multiple} express the propagator entirely in terms of the low-dimensional system operators, the time evolution of these operators requires computing the time evolution of the entire system which remains computationally intractable due to the high dimensionality of the Hilbert space of the waveguide modes. In this section, we show that these Green's functions can equivalently be computed by evolving the low dimensional system under an effective non-hermitian Hamiltonian. We consider a general Green's function of the form:
\begin{equation}\label{eq:gen_green_func}
\mathcal{G}_{\tau_+, \tau_-}^{\sigma',\sigma}(t_1, t_2 \dots t_N) = \bra{\text{vac};\sigma' }U_\textrm{I}(\tau_+,0)\mathcal{T}\big[s_1(t_1) s_2(t_2) \dots s_N(t_N)\big] U_\textrm{I}(0,\tau_-)\ket{\text{vac};\sigma }
\end{equation}
where $s_i$ are operators defined in the low-dimensional system's Hilbert space and $\tau_-<t_i<\tau_+ \ \forall \ i\in\{1,2\dots N\}$. Let $P$ be a permutation of $\{1,2,3 \dots N\}$ such that $t_{P(1)}\geq t_{P(2)} \dots \geq t_{P(N)}$, then
\begin{equation}
\mathcal{G}^{\sigma', \sigma}_{\tau_+,\tau_-}(t_1, t_2 \dots t_N) = \langle \text{vac}; \sigma' | U_\textrm{I}(\tau_+,0)\bigg[ \prod_{i=1}^N s_{P(i)}(t_{P(i)})\bigg] U_\textrm{I}(0,\tau_-) |\text{vac}; \sigma \rangle.
\end{equation}
We note that the system operators $s_i$ commute with the waveguide Hamiltonian $H_\text{wg}$ and thus $s_i(t) = U(0,t) s_i U(t,0) = U(0,t) \exp(-\textrm{i}H_\text{wg}t) s_i\exp(\textrm{i}H_\text{wg}t) U(t,0) = U_\textrm{I}(0,t) s_i U_\textrm{I}(t,0)$. The Green's function can thus be expressed as:
\begin{align}\label{green_func_schro}
\mathcal{G}_{\tau_-, \tau_+}^{\sigma',\sigma}(t_1, t_2 \dots t_N) = \langle \text{vac}; \sigma' | U_\textrm{I}(\tau_+,t_{P(1)}) \bigg[\prod_{i=1}^{N-1} s_{P(i)}U_\textrm{I}(t_{P(i)},t_{P(i+1)}) \bigg] s_{P(N)}U_\textrm{I}(t_{P(N)},\tau_-)  |\text{vac}; \sigma \rangle.
\end{align}

We next show that the vacuum expectation in this equation can be explicitly evaluated --- as a starting point, we express the interaction picture propagator in terms of the interaction picture Hamiltonian:
\begin{align}
U_\textrm{I}(t_1, t_2) = \mathcal{T}\exp\bigg[-\textrm{i}\int_{t_2}^{t_1} H_\textrm{I}(t') \textrm{d}t' \bigg]
\end{align}
where
\begin{align}\label{eq:int_hamil}
H_\textrm{I}(t) = H_\text{sys}(t)+\sum_{\mu=1}^{N_L} \int \big( a_{\omega, \mu} L_\mu^\dagger \exp(-\textrm{i}\omega t)+L_\mu a_{\omega, \mu}^\dagger \exp(\textrm{i}\omega t) \big) \frac{\textrm{d}\omega}{\sqrt{2\pi}}.
\end{align}
In terms of the spatial annihilation operator, $a_{x,\mu}$, the interaction picture Hamiltonian can be rewritten as:
\begin{align}
H_\textrm{I}(t) = H_\text{sys}(t) + \sum_{\mu=1}^{N_L}\big(a_{x = -t,\mu} L_\mu^\dagger+L_\mu a_{x = -t,\mu}^\dagger \big).
\end{align}
To proceed, we approximate the high-dimensional continuum Hilbert space of the waveguides by a discrete Hilbert Space with a countably infinite basis (Fig.~\ref{fig:fig_apprx}). This is achieved by introducing a coarse graining parameter $\delta x$, and defining the `coarse grained operators' $A_{\mu, n}$ via:
\begin{align}
A_{\mu}[n] = \int\limits_{(n-1)\delta x}^{n\delta x}a_{x,\mu}\frac{\textrm{d}x}{\sqrt{\delta x}}.
\end{align}
These operators satisfy the commutators $[A_{\mu}[n], A_{\mu'}^\dagger[n']]= \delta_{\mu,\mu'}\delta_{n,n'}, [A_{\mu}[n], A_{\mu'}[n']] = 0$ and in the limit of $\delta x \to 0$, $A_\mu[n]/\sqrt{\delta x}$ would approach the continuum operator $a_{x = n\delta x,\mu}$. The discrete Hilbert space approximating the waveguide's Hilbert space would then be the space spanned by the tensor product of the Fock states created by each of the operators $A_\mu^\dagger[n], \ \forall \mu \in \{1, 2\dots N_L\}$ and $n \in \mathbb{Z}$. In particular, the vacuum state $\ket{\text{vac}}$ for the Hilbert space of the waveguide is approximated by
\begin{equation}
\bigotimes_{\mu = 1}^{N_L} \bigotimes_{n=-\infty}^{\infty} \ket{\text{vac}_{n,\mu}}
\end{equation}
where $\ket{\text{vac}_{n,\mu}}$ is the vacuum state corresponding to the operator $A_\mu[n]$. Moreover, this interaction Hamiltonian (Eq.~\ref{eq:int_hamil}) is then expressed as $H_\textrm{I}(t; \delta x)$ in the limit of $\delta x \to 0$, where $H_\textrm{I}(t; \delta x)$ is given by: 
\begin{align}\label{eq:disc_hamil}
H_\textrm{I}(t; \delta x) = H_\text{sys}(t) + \frac{1}{\sqrt{\delta x}}\bigg\{\sum_{\mu=1}^{N_L} \big(A_\mu\big[\lceil -t/\delta x \rceil\big] L_\mu^\dagger+L_\mu A_\mu^\dagger \big[\lceil -t/\delta x \rceil\big] \big) \bigg \} .
\end{align}
Using the notation $n_+ = \lceil \tau_+/\delta x\rceil$, $n_- = \lceil \tau_-/\delta x \rceil$, $n_i = \lceil t_i/\delta x \rceil$ and defining
\begin{align}
U_\textrm{I}[n_1, n_2] = \mathcal{T}\exp\bigg(-\textrm{i}\int \limits_{n_1 \delta x}^{n_2 \delta x}H_\textrm{I}(t; \delta x) \textrm{d}x \bigg)
\end{align}
the Green's function in Eq.~\ref{green_func_schro} can be expressed as:
\begin{align}\label{eq:green_func_disc}
\mathcal{G}^{\sigma', \sigma}_{\tau_+,\tau_-}(t_1, t_2 \dots t_N) & = \lim_{\delta x \to 0} \bigg[\bigg\{\bigotimes_{\mu = 1}^{N_L} \bigotimes_{n=-\infty}^{\infty}  \bra{\text{vac}_{n,\mu}} \bigg\}\otimes \bra{\sigma'}\bigg]U_\textrm{I}[n_+, n_{P(1)}]\bigg[\prod_{i=1}^{N-1} s_{P(i)}U_\textrm{I}[n_{P(i)}, n_{P(i+1)}]\bigg]\times \nonumber \\
& \qquad s_{P(N)}U_\textrm{I}[n_{P(N)}, n_-] \bigg[\bigg\{\bigotimes_{\mu = 1}^{N_L} \bigotimes_{n=-\infty}^{\infty} \ket{\text{vac}_{n,\mu}}\bigg\}\otimes \ket{\sigma} \bigg].
\end{align}
Noting that since for $n\delta x < t <(n+1) \delta x$, $H_\textrm{I}(t)$ depends only on $A_{\mu}[-n]$, $U_\textrm{I}[n+1, n]$ would only act on $\ket{\text{vac}_{-n,\mu}} \ \forall \mu$. Using this together with the decomposition $U_\textrm{I}[n_1, n_2] = U_\textrm{I}[n_1,n_1-1]U_\textrm{I}[n_1-1, n_1-2] \dots U_\textrm{I}[n_2-1, n_2]$, we can rewrite Eq.~\ref{eq:green_func_disc} as:
\begin{align}\label{eq:gfunc_eff_single_step}
\mathcal{G}_{\tau_+, \tau_-}^{\sigma',\sigma}(t_1, t_2 \dots t_N) & = \lim_{\delta t \to 0} \bra{\sigma'} \bigg[ \prod_{j = -n_+-1}^{-n_{P(1)}}U_\textrm{eff}[-j+1,-j] \bigg] \bigg[\prod_{i=1}^{N-1}\bigg\{s_{P(i)}\prod_{j=-n_{P(i)}-1}^{-n_{P(i+1)}} U_\text{eff}[-j+1,-j] \bigg\}\bigg]\times \nonumber \\ &  \qquad s_{P(N)}\bigg[ \prod_{j=-n_{P(N)}-1}^{-n_{-}}U_\text{eff}[-j+1,-j]\bigg]\ket{\sigma}
\end{align}
where
\begin{align}
U_\text{eff}[n+1,n] = \bigg[\bigotimes_{\mu=1}^{N_L}\bra{\text{vac}_{-n,\mu}}\bigg] U_\textrm{I}[n+1,n]\bigg[\bigotimes_{\mu=1}^{N_L}\ket{\text{vac}_{-n,\mu}}\bigg].
\end{align}
This expression can be further simplified by expanding $U_\textrm{I}[n+1,n]$ into a Dyson series in terms of the interaction Hamiltonian:
\begin{align}
U_\textrm{I}[n+1,n] = \textrm{I}+\sum_{n=1}^{\infty}(-\textrm{i})^n \int\limits_{t_1 = n\delta x}^{(n+1)\delta x}\int\limits_{t_2 = t_1}^{(n+1)\delta x} \dots \int\limits_{t_n =t_{n-1}}^{(n+1)\delta x} H_\textrm{I}(t_n; \delta x)H_\textrm{I}(t_{n-1}; \delta x) \dots H_\textrm{I}(t_1; \delta x)  \textrm{d}t_1 \dots \textrm{d}t_n .
\end{align}

It follows from Eq.~\ref{eq:disc_hamil} that the vacuum expectations corresponding to the first two terms in the summation in the Dyson series evaluate to:
\begin{subequations}
\begin{align}
&\bigg[\bigotimes_{\mu=1}^{N_L}\bra{\text{vac}_{-n,\mu}}\bigg] \int\limits_{t = n\delta x}^{(n+1)\delta x}H_\textrm{I}(t'; \delta x) \textrm{d}t' \bigg[\bigotimes_{\mu=1}^{N_L}\ket{\text{vac}_{-n,\mu}}\bigg] = \int\limits_{n\delta x}^{(n+1)\delta x}H_\text{sys}(t) \textrm{d}t \\
&\bigg[\bigotimes_{\mu=1}^{N_L}\bra{\text{vac}_{-n,\mu}}\bigg] \int\limits_{t_1 = n\delta x}^{(n+1)\delta x} \int\limits_{t_2 = t_1} ^{(n+1)\delta x} H_\textrm{I}(t_2; \delta x) H_\textrm{I}(t_1; \delta x) \textrm{d}t_1 \textrm{d}t_2 \bigg[\bigotimes_{\mu=1}^{N_L}\ket{\text{vac}_{-n,\mu}}\bigg] = 
\frac{1}{2} \sum_{\mu=1}^{N_L} L_\mu^\dagger L_\mu \delta x + \mathcal{O}(\delta x^2).
\end{align}
\end{subequations}
Moreover, it can easily be seen from Eq.~\ref{eq:int_hamil} that the vacuum expectations of higher order terms in the Dyson series do not have any contributions that are first order in $\delta x$. Therefore,
\begin{align}\label{eq:eff_single_step}
U_\textrm{eff}[n+1, n] = \textrm{I}-\textrm{i}\int\limits_{t=n\delta x}^{(n+1)\delta x}H_\text{eff}(t') \textrm{d}t' +\mathcal{O}(\delta x^2) = \mathcal{T} \exp \bigg[-\textrm{i}\int \limits_{n\delta x}^{(n+1)\delta x}H_\text{eff}(t) \textrm{d}t \bigg] + \mathcal{O}(\delta x^2)
\end{align}
where 
\begin{align}
H_\text{eff}(t) = H_\text{sys}(t)-\frac{\textrm{i}}{2}\sum_\mu L_\mu^\dagger L_\mu.
\end{align}
Finally, substituting Eq.~\ref{eq:eff_single_step} into Eq.~\ref{eq:gfunc_eff_single_step} and evaluating the limit, we obtain:
\begin{align}\label{eq:greens_func_eff}
\mathcal{G}_{\tau_-, \tau_+}^{\sigma',\sigma}(t_1, t_2 \dots t_N) = \langle  \sigma' | U_\textrm{eff}(\tau_+,t_{P(1)}) \bigg[\prod_{i=1}^{N-1} s_{P(i)}U_\textrm{eff}(t_{P(i)},t_{P(i+1)}) \bigg] s_{P(N)}U_\textrm{eff}(t_{P(N)},\tau_-)  | \sigma \rangle
\end{align}
where
\begin{align}
U_\text{eff}(t_1, t_2) = \mathcal{T}\exp\bigg[-\textrm{i}\int\limits_{t_2}^{t_1}H_\text{eff}(t) \textrm{d}t \bigg].
\end{align}
Eq.~\ref{eq:greens_func_eff} is an expectation evaluated entirely in the Hilbert space of the low-dimensional system, which makes it computationally tractable. For many systems of interest, it is often easier to work with a Heisenberg-like form of the Green's function, which can be obtained by defining $\tilde{s}_i(t) = U_\text{eff}(0,t)s_i U_\text{eff}(t,0)$, which satisfies the Heisenberg-like equations of motion:
\begin{align}
\dot{\tilde{s}}_i(t) = -\textrm{i}[\tilde{H}_\text{eff}(t), \tilde{s}_i(t)]
\end{align}
where $\tilde{H}_\text{eff}(t) = U_\text{eff}(0,t) H_\text{eff}(t) U_\text{eff}(t,0)$. In terms of these operators, Eq.~\ref{eq:greens_func_eff} can be reduced to
\begin{align}
\mathcal{G}_{\tau_+,\tau_-}^{\sigma',\sigma}(t_1, t_2 \dots t_N) &= \langle \sigma' |U_\text{eff}(\tau_+,0) \bigg[\prod_{i=1}^N \tilde{s}_P(i)(t_{P(i)}) \bigg] U_\text{eff}(0,\tau_{-}) \ket{\sigma}  \nonumber \\ &= \langle \sigma' |U_\text{eff}(\tau_+,0) \mathcal{T}\bigg[\prod_{i=1}^N \tilde{s}_i(t_{i}) \bigg] U_\text{eff}(0,\tau_{-}) \ket{\sigma} .
\end{align}

\section{\label{sec:scat_matrix}Scattering matrices}
\noindent The scattering matrix is a useful quantity to characterize the response of the low dimensional system to wave-packets incident from the waveguide modes coupling to it. The scattering matrix $\Sigma$ can be computed from the interaction picture operator by taking the limits $\tau_{-} \to -\infty$ and $\tau_{+} \to \infty$:
\begin{align}
\Sigma = \lim_{\substack{\tau_+ \to \infty \\ \tau_- \to -\infty}} U_I(\tau_+,\tau_-).
\end{align}
The scattering matrix can be completely characterized by matrix elements of the form:
\begin{align}
&\Sigma^{\sigma', \sigma}(\{x_1',\mu_1'\}\dots \{x_M',\mu_M'\}; \{x_1,\mu_1\} \dots \{x_N,\mu_N\}) \nonumber \\ &=\bra{\{x_1',\mu_1'\}\dots \{x_M',\mu_M'\};\sigma'}\Sigma \ket{ \{x_1,\mu_1\} \dots \{x_N,\mu_N\};\sigma}.
\end{align}
In order to compute the scattering matrix from the propagator analyzed in the previous section, we make the definition of the system Hamiltonian $H_\text{sys}(t)$ more explicit --- in particular, we assume that it is `time dependent' only if $t \in [0,T_P]$. Physically, this might correspond to a system like a two-level atom being driven by a coherent pulse which vanishes outside $[0,T_P]$. Mathematically, this is equivalent to writing $H_\text{sys}(t)$ as:
\begin{align}
H_\text{sys}(t) = 
\begin{cases}
H_\text{sys}^{0}+H_\text{sys}^P(t) & \text{for} \ 0\leq t\leq T_P \\
H_\text{sys}^{0} & \text{otherwise}
\end{cases}.
\end{align}

For a low-dimensional system Hamiltonian of this form and by using Eq.~\ref{eq:prop_green_func} the scattering matrix element can be expressed as:
\begin{align}
&\Sigma_{\sigma',\sigma}(\{x_1',\mu_1'\}\dots \{x_M',\mu_M'\}; \{x_1,\mu_1\} \dots \{x_N,\mu_N\}) =\sum_{k=0}^N (-\textrm{i})^{M+N-2k}\mathcal{I}(M\geq k) \times  \\ &\quad\sum_{B_k^N, B_k^M} \bigg[\sum_{P_k} \prod_{i=1}^k \delta(x'_{P_k B_k^M(i)}-x_{B_k^N(i)})\delta_{\mu_{P_k B_k^M(i)}', \mu_{B_k^N(i)}} \bigg] \times \nonumber \\  
&\quad \mathcal{G}^{\sigma',\sigma}_{\infty,-\infty}(\{-x'_{\bar{B}_k^M(1)}, \mu_{\bar{B}_k^M(1)}'\}, \dots \{-x'_{\bar{B}_k^M(M-k)},\mu'_{\bar{B}_k^M(M-k)}\};\{-x_{\bar{B}_k^N(1)},\mu_{\bar{B}_k^N(1)}\} \dots \{-x_{\bar{B}_k^N(N-k)},\mu_{\bar{B}_k^N(N-k)}\})\nonumber
\end{align}
where
\begin{align}
&\mathcal{G}_{\infty,-\infty}^{\sigma',\sigma}(\{t_1',\mu_1'\} \dots \{t_M',\mu_M'\}; \{t_1,\mu_1\} \dots \{t_N,\mu_N\}) \nonumber \\
&=\bra{\sigma'}U_\textrm{eff}^0(\tau_+,T_P)U_\text{eff}(T_P,0)\mathcal{T}\bigg[\prod_{i=1}^M \tilde{L}_{\mu_i'}(t_i') \prod_{j=1}^N \widetilde{L^\dagger}_{\mu_j}(t_j) \bigg]U_\textrm{eff}^0(0,\tau_-)\ket{ \sigma}
\end{align}
with
\begin{align}
U_\text{eff}^0(t_1,t_2) = \exp\bigg[-\bigg(\textrm{i}H_\text{sys}^0+\sum_\mu \frac{1}{2} L_\mu^\dagger L_\mu \bigg)(t_1-t_2)\bigg].
\end{align}

While the orthonormal basis $\{\ket{\sigma_1}, \ket{\sigma_2} \dots \ket{\sigma_S}\}$ for the low-dimensional system's Hilbert space used for computing the propagator could be arbitrarily chosen as long as it is complete, for the purpose of computing the scattering matrix, we make this basis more explicit by expressing it as a union of a set of `ground states' $\{\ket{g_1}, \ket{g_2} \dots \ket{g_{S_g}}\}$ and `excited states' $\{\ket{e_1}, \ket{e_2} \dots \ket{e_{S_e}}\}$ which are all eigen-states of the Hamiltonian $H_\text{sys}^0$:
\begin{align}
H_\text{sys}^0\ket{g_n} = \varepsilon_{g_n}\ket{g_n} \ \text{and} \ H_\text{sys}^0 \ket{e_n} = \varepsilon_{e_n}\ket{e_n}.
\end{align}
Moreover, the ground states and the excited states also satisfy:
\begin{subequations}\label{eq:def_states}
\begin{align}
&L_\mu \ket{g_n} = 0 \ \forall \ \mu \in \{1,2 \dots N_L\}, n \in \{1, 2 \dots S_g\}, m \in \{1,2 \dots S_e\}  \\
&\exists \mu \in \{1,2 \dots N_L\} \ |\  L_\mu \ket{e_n} \neq 0 \ \forall \ n \in \{1,2 \dots S_e \} \\
& \braket{e_m | g_n} = 0 \ \forall \ m \in \{1,2 \dots S_e \}, n \in \{1,2 \dots S_g \} 
\end{align}
\end{subequations}
where $L_\mu$ are the operators through which the low dimensional system couples to the waveguide modes. An immediate consequence of the definition of ground and excited states (Eq.~\ref{eq:def_states}) and the positive-definiteness of the operator $\sum_\mu L_\mu^\dagger L_\mu$ (which is also the non-hermitian part of the effective Hamiltonian) within the subspace spanned by the excited states, is that if evolved with the propagator $U_\text{eff}^0(t_1,t_2)$ an excited state would decay to 0 and a ground state would remain unchanged except for accumulating a phase: 
\begin{subequations}
\begin{align}
&\lim_{\tau_- \to -\infty} U_\text{eff}^0(0,\tau_-)|e_n\rangle = 0 \ \text{and} \ \lim_{\tau_- \to -\infty} U_\text{eff}^0(0,\tau_-)|g_n\rangle = \exp(\textrm{i}\varepsilon_{g_n} \tau_-)\ket{g_n} \\
&\lim_{\tau_+ \to \infty} \bra{e_n}U_\text{eff}^0(\tau_+,T_P) = 0 \ \text{and} \ \lim_{\tau_+ \to -\infty} \bra{g_n}U_\text{eff}^0(\tau_+,T_P) = \exp(-\textrm{i}\varepsilon_{g_n} (\tau_+-T_P))\bra{g_n}.
\end{align}
\end{subequations}
Therefore, only the scattering matrix elements corresponding to the low-dimensional system going from one ground state to another ground state are non zero and are given by:
\begin{align}\label{eq:scat_matrix_final}
&\Sigma_{g_m,g_n}(\{x_1',\mu_1'\}\dots \{x_M',\mu_M'\}; \{x_1,\mu_1\} \dots \{x_N,\mu_N\}) =\sum_{k=0}^N (-\textrm{i})^{M+N-2k}\mathcal{I}(M\geq k) \times  \\ &\quad\sum_{B_k^N, B_k^M} \bigg[\sum_{P_k} \prod_{i=1}^k \delta(x'_{P_k B_k^M(i)}-x_{B_k^N(i)})\delta_{\mu_{P_k B_k^M(i)}', \mu_{B_k^N(i)}} \bigg] \times \nonumber \\  
&\quad \mathcal{G}^{\sigma',\sigma}_{\infty,-\infty}(\{-x'_{\bar{B}_k^M(1)}, \mu_{\bar{B}_k^M(1)}'\}, \dots \{-x'_{\bar{B}_k^M(M-k)},\mu'_{\bar{B}_k^M(M-k)}\};\{-x_{\bar{B}_k^N(1)},\mu_{\bar{B}_k^N(1)}\} \dots \{-x_{\bar{B}_k^N(N-k)},\mu_{\bar{B}_k^N(N-k)}\})\nonumber
\end{align}
where
\begin{align}\label{eq:scat_matrix_gfunc}
&\mathcal{G}_{\infty,-\infty}^{g_m ,g_n}(\{t_1',\mu_1'\} \dots \{t_M',\mu_M'\}; \{t_1,\mu_1\} \dots \{t_N,\mu_N\}) =\bra{g_m}U_\text{eff}(T_P,0)\mathcal{T}\bigg[\prod_{i=1}^M \tilde{L}_{\mu_i'}(t_i') \prod_{j=1}^N \widetilde{L^\dagger}_{\mu_j}(t_j) \bigg]\ket{g_n}
\end{align}
wherein we have dropped the phase factors depending on $\tau_+$ and $\tau_-$ corresponding to the phase accumulated by the ground states from the scattering matrix element. Previous calculation of the scattering matrix \cite{xu2015input,xu2017input} for systems with time-independent Hamiltonians arrived at exactly the same form as in Eqs.~\ref{eq:scat_matrix_final} and \ref{eq:scat_matrix_gfunc} with $T_P = 0$. However, as previously emphasized, the formalism introduced here allows us to model systems with time-dependent Hamiltonians, thereby increasing its applicability in modeling experimentally relevant systems.

\section{\label{sec_examples} Examples}
{In this section, we show how to use the formalism developed in the previous sections to analyze some open-quantum systems of interest. The examples we choose to analyze include a two-level system and a lambda three-level system. We calculate both emission from these systems when they are coherently driven, and scattering of single-photon pulses from these systems.}
\subsection{\label{sec:tls}{Two-level System}}
\begin{figure}
\centering
\includegraphics[scale = 0.9]{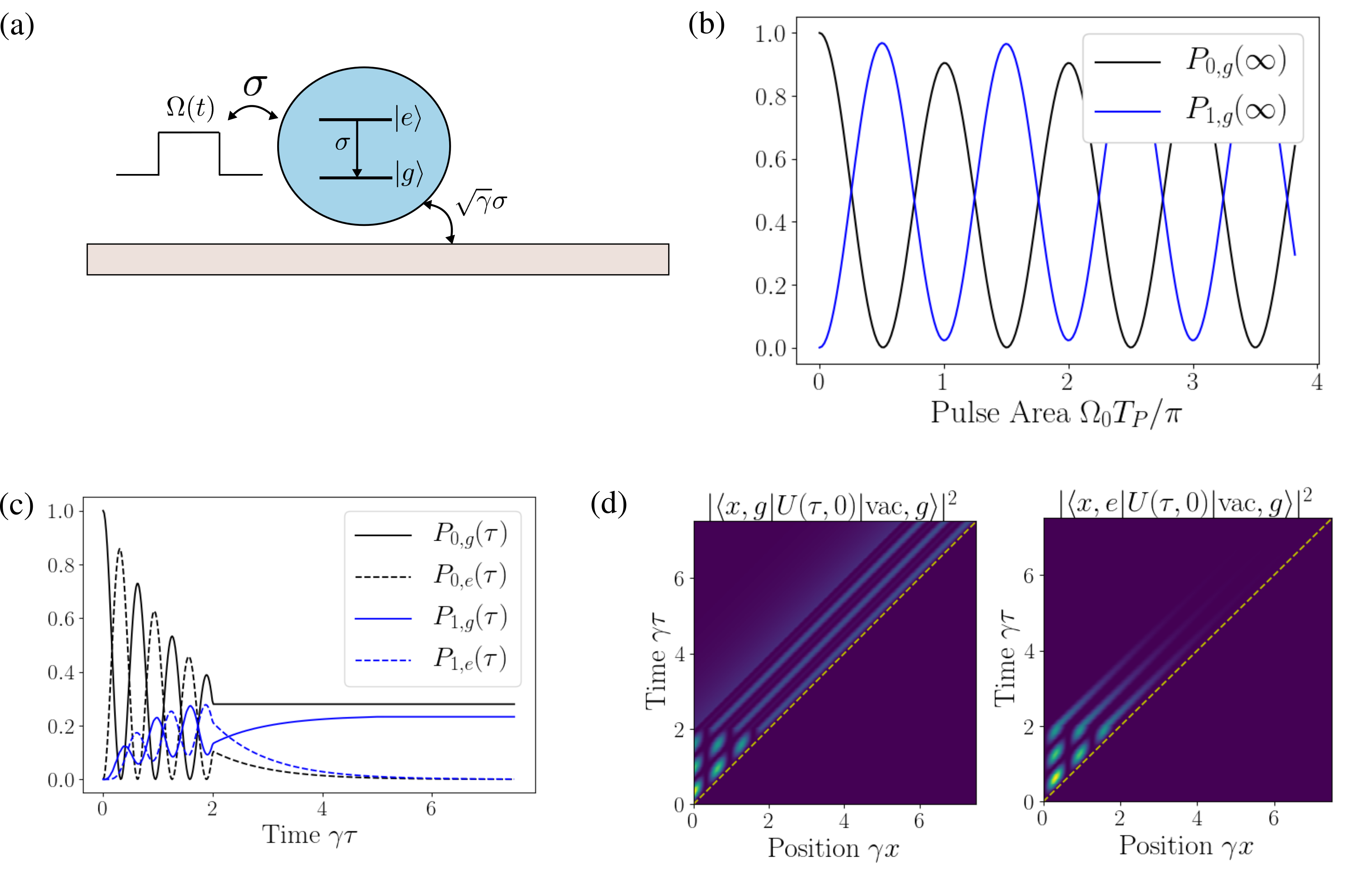}
\caption{Emission from a coherently driven two-level system. (a) Schematic of a two-level system driven with a pulse $\Omega(t)$. (b) Probability of emission of a single photon $P_{0,g}(\infty)$ and of no emission $P_{1,g}(\infty)$ from a two-level system driven by a short pulse ($\gamma T_P = 0.2$) as a function of the pulse area. (c) Time dependence of the probabilities $P_{0,g}, P_{1,g}, P_{0,e}$ and $P_{1,e}$ for a two-level system driven by a long pulse ($\gamma T_P = 2$ and $\Omega_0 = 5\gamma $). (d) Space-time dependence of the propagator matrix elements corresponding to a single photon in the waveguide (dotted line is the light line $x=\tau$).}
\label{fig:tls_emission}
\end{figure}

\noindent As our first example, we consider a coherently driven two-level system  coupled to a single waveguide with coupling decay rate $\gamma$. Denoting the ground state of the two-level system with $\ket{g}$ and the excited state with $\ket{e}$, the two-level system can be modeled with the following Hamiltonian:
\begin{align}\label{eq:tls_hamiltonian}
H_\text{sys} = \delta_a \sigma^\dagger \sigma + \Omega(t)(\sigma+\sigma^\dagger)
\end{align}
where $\sigma = \ket{g}\bra{e}$ and $\sigma^\dagger = \ket{e}\bra{g}$ are the annihilation and creation operators for the two-level system, $\delta_a$ is the detuning of the resonant frequency of the two level atom from the frequency of the coherent drive and $\Omega(t)$ is the amplitude of the coherent drive, which we assume to be of the form:
\begin{align}\label{eq:coh_drive}
\Omega(t) = 
\begin{cases}
\Omega_0 & 0 \leq t \leq T_P \\
0 & \text{otherwise}
\end{cases}.
\end{align}
We first analyze emission from this driven two-level system --- we consider the two-level system coupled to a single waveguide, with it being in the ground state and the waveguide to be in the vacuum state at $t = 0$ [Fig.~\ref{fig:tls_emission}(a)]. This amounts to computing the propagator $U_I(\tau,0)$, from which it is easy to extract the state of the waveguide and the two-level system as a function of $\tau$. Of particular interest are the probabilities of finding 0 and 1 photon in the waveguide with the two-level system being in the excited or ground state as a function of $\tau$:
\begin{subequations}\label{eq:tls_prob}
\begin{align}
&P_{0,g}(\tau) = |\langle \text{vac}; g| U(\tau,0)\ket{\text{vac}; g}|^2, \quad\ P_{0,e}(\tau) = |\langle \text{vac}; e|  U(\tau,0)\ket{\text{vac}; g}|^2 \\
&P_{1,g}(\tau) = \int |\langle x; g| U(\tau,0)\ket{\text{vac};g}|^2 \text{d}x; \quad P_{1,e}(\tau) = \int |\langle x; e| U(\tau,0)\ket{\text{vac};g}|^2 \text{d}x .  
\end{align}
\end{subequations}
Fig.~\ref{fig:tls_emission}(b) shows the steady state probabilities $P_{1,g}(\infty)$ and $P_{0,g}(\infty)$ of a two-level system emitting a single photon or not emitting any photons after being excited by a short pulse as  a function of the pulse area. We observe the well understood rabi-oscillations in these probabilities with the pulse area. Fig.~\ref{fig:tls_emission}(c) shows the time-dependence of the probabilities defined in Eq.~\ref{eq:tls_prob} for a long pulse --- again, we observe oscillation in these probabilities while the two-level system is being driven, followed by a decay of the two-level system to its ground state and emission of photons into the waveguide. The full space-time dependence of the propagator matrix elements corresponding to a single photon in the waveguide is shown in Fig.~\ref{fig:tls_emission}(d) --- we clearly see signatures of rabi-oscillation during the time interval in which the two-level system is being driven by the coherent pulse. During this time interval, the two-level system state and the waveguide state are entangled to each other, and after the coherent pulse goes to 0 the two-level system completely decays into the waveguide mode and the resulting excitation propagates along the waveguide.  It can also be noted that the matrix elements are always zero outside the light-cone (i.e. for $x>t$), which is intuitively expected since the group velocity is an upper bound on the speed at which photons emitted by the two-level system can propagate in the waveguide.

\begin{figure}
\includegraphics[scale = 0.8]{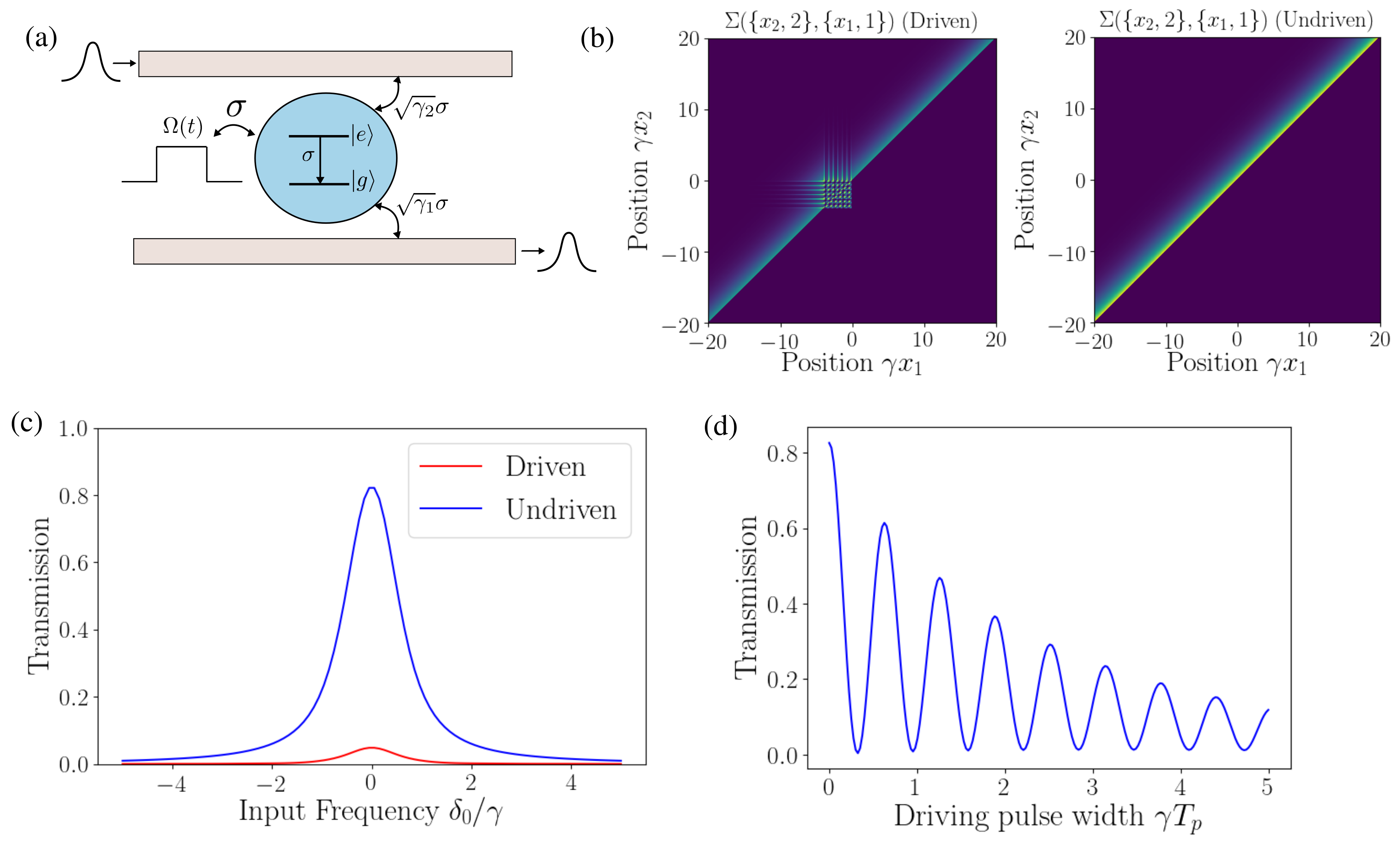}
\caption{Scattering of a single-photon wave-packet from a coherently driven two-level system. (a) Schematic of a coherently driven two-level system coupled to an input and output waveguide. (b) Single-photon scattering matrix for the driven and undriven two-level system. (c) Variation of the transmission of the coherently driven two-level system with the central frequency $\delta_0$ of the input single-photon state. (d) Variation of the transmission of a coherently driven two-level system with the length of the coherent drive for a resonant input single photon state ($\delta = 0$). For (b) and (c), it is assumed that $\gamma T_P = 4$ and $\Omega_0 = 5\gamma$. $\gamma_1 = \gamma_2 = \gamma/2$ and $\Delta x = 2.0/\gamma$ is assumed in all the calculations.}
\label{fig:tls_scattering}
\end{figure} 

Next, we analyze scattering of a single-photon pulse from a coherently driven two-level system (Fig.~\ref{fig:tls_scattering}) --- this amounts to computing the scattering matrix for the time dependent Hamiltonian in Eq.~\ref{eq:tls_hamiltonian} using Eq.~\ref{eq:scat_matrix_final}. We consider a two-level system coupled to two waveguides (Fig.~\ref{fig:tls_scattering}), and excite it with an input pulse from the first waveguide (labeled as 1), and compute the single-photon component of the output state in the second waveguide (labeled as 2). Fig.~\ref{fig:tls_scattering}(b) shows the single-photon scattering matrix for a two-level system, with and without the coherent drive. The two scattering matrices differ in the region $-T_P < x_1,x_2<0$, which corresponds to incident pulses that arrive at the two-level system while it is being driven.  To analyze the transmission spectra of the driven two-level system, we consider exciting the first waveguide with a single-photon state (in the interaction picture) at time $-\infty$ of the form:
\begin{align}
\ket{\psi(-\infty)} = \int \psi_\text{in}(x_1)  a_{x_1,1}^\dagger \ket{\text{vac}} \textrm{d}x_1, \quad \psi_\text{in}(x_1) = \frac{1}{(\pi \Delta x^2)^{1/4}}\exp\bigg(-\frac{x_1^2}{2\Delta x^2}+\textrm{i}\delta_0 x_1 \bigg)
\end{align}
where $\delta_0$ is the central frequency of the incident wave-packet, and $\Delta x$ is its spatial extent. We compute the single-photon component in the output waveguide by applying the scattering matrix to the input state:
\begin{align}
\psi_\text{out}(x_2) = \int \Sigma(\{x_2, 2\}, \{x_1, 1\}) \psi_\text{in}(x_1) \textrm{d}x_1
\end{align}
and the transmission of the single-photon state via:
\begin{align}
\text{Transmission} = \int |\psi_\text{out}(x_2)|^2 \textrm{d}x_2.
\end{align}
Figure \ref{fig:tls_scattering}(c) shows the transmission spectrum of the two-level system with and without the coherent drive --- the driven two-level system clearly shows a suppression in transmission. Intuitively, this is a consequence of the coherent pulse transferring the two-level system into its excited state when the single-photon pulse arrives at the two-level system, thereby resulting in low transmission into the output waveguide. Figure~\ref{fig:tls_scattering} shows the dependence of the transmission of the two-level system for a resonant input single-photon wave-packet ($\delta_0 = 0$) on the length of the coherent drive $T_P$. Again, we see a signature of the rabi-oscillation in the transmission --- the probability of the two-level system being in the excited state oscillates with the input pulse, and this oscillation translates to the oscillation in the transmission of the driven two-level system. 

\subsection{\label{lambda}{Lambda system}}
\noindent As our next example, we consider a coherently driven three-level lambda-system. This system has two ground states, denoted by $\ket{g_1}$ and $\ket{g_2}$, and one excited state, denoted by $\ket{e}$. The Hamiltonian for this system is given by:
\begin{align}
H_\text{sys} = \delta_e \ket{e}\bra{e}+\delta_{12}\ket{g_1}\bra{g_1}+\Omega(t)(\sigma_1+\sigma_1^\dagger)
\end{align}
where $\sigma_i = \ket{g_i}\bra{e}$ is the operator that annihilates the excited state $\ket{e}$ to the ground state $\ket{g_i}$, $\delta_e$ is the detuning of the frequency difference between $\ket{e}$ and $\ket{g_2}$ from the frequency of the coherent drive, $\delta_{12}$ is the frequency difference between the states $\ket{g_2}$ and $\ket{g_1}$, and $\Omega(t)$ is the amplitude of the coherent drive which we again assume to be a rectangular pulse as given by Eq.~\ref{eq:coh_drive}. 
\begin{figure}
\centering
\includegraphics[scale = 0.9]{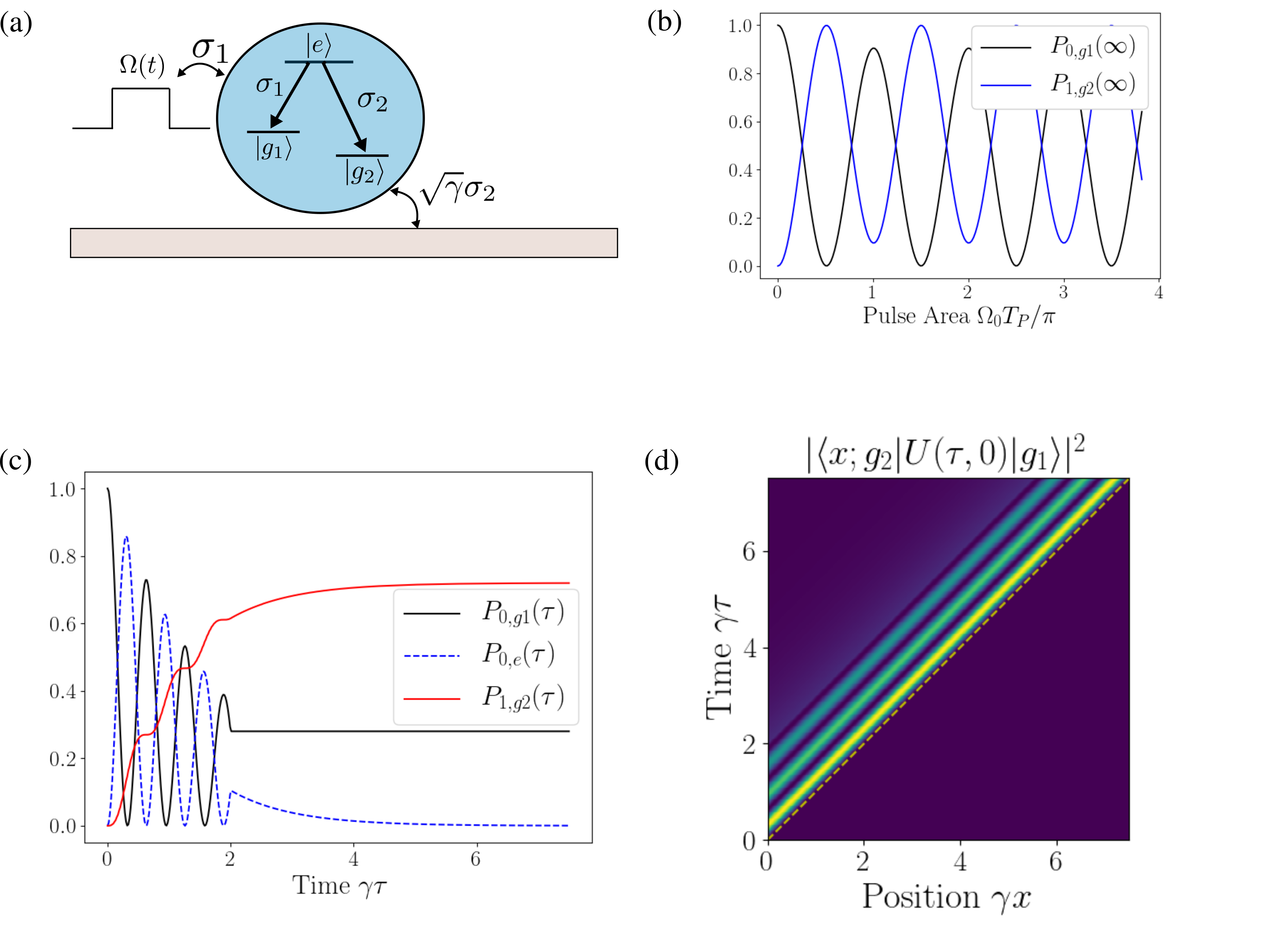}
\caption{Emission from a coherently driven three-level lambda system. (a) Schematic of the system being analyzed. (b) Probability of emission of a single photon $P_{1,g_2}(\infty)$ and of no emission $P_{0,g_1}(\infty)$ from the lambda system driven by a short pulse ($\gamma T_P = 0.2$) as a function of the pulse area. (c) Time dependence of the probabilities $P_{0,g_1}, P_{1,g_2}$ and $P_{0,e}$ for a three-level lambda system driven by a long pulse ($\gamma T_P = 2$ and $\Omega_0 = 5\gamma$). (d) Space-time dependence of the propagator matrix elements corresponding to a single photon in the waveguide (dotted line is the light line $x=\tau$). $\delta_{12} = 0$ and $\delta_e = 0$ are assumed in all the simulations.}
\label{fig:lambda_1}
\end{figure}

We first analyze photon emission from this driven lambda system --- the lambda system is assumed to couple to a single waveguide through an operator $\sqrt{\gamma} \sigma_2$, with it being in the ground state $\ket{g_1}$ and the waveguide being in the vacuum state at time 0 [Fig.~\ref{fig:lambda_1}(a)]. The probabilities of interest that we compute include the following:
\begin{align}\label{eq:lambda_sys_prob}
&P_{0,g1}(\tau) = |\bra{\text{vac}; g_1}U_\textrm{I}(\tau,0) \ket{\text{vac}; g_1}|^2, P_{0, e}(\tau) = |\bra{\text{vac}; e}U_\textrm{I}(\tau,0)\ket{\text{vac}; g_1}|^2 \nonumber \\
&P_{1,g2} = \int |\bra{x; g_2} U_\textrm{I}(\tau,0) \ket{\text{vac}; g_1}|^2 \textrm{d}x
\end{align}
Figure~\ref{fig:lambda_1}(b) shows the dependence of the probabilities of single photon emission and no emission as a function of the pulse area for a short driving pulse -- we again observe the expected rabi-oscillations in these probabilities. Fig.~\ref{fig:lambda_1}(c) shows the time evolution of the probabilities defined in Eq.~\ref{eq:lambda_sys_prob} for a long driving pulse --- we clearly observe an oscillation in these probabilities while the lambda system is being driven followed by emission of a single photon into the waveguide. We note that for a lambda system, once a photon emission into the waveguide occurs, the lambda system would necessarily transition to the state $\ket{g_2}$, and will no longer be driven by $\Omega(t)$. As a consequence of this structure of the lambda system, only a single photon can be emitted into the waveguide --- this is numerically validated in Fig.~\ref{fig:lambda_1}(c) from which it can be seen that $P_{0,g1}(\infty)+P_{1, g2}(\infty) = 1$. Figure~\ref{fig:lambda_1}(d) shows the propagator matrix element $|\bra{x;g_2}U(\tau,0)\ket{\text{vac};g_1}|^2$ --- we clearly see a stark difference from the corresponding matrix element for a two-level system (Fig.~\ref{fig:tls_emission}) due to the system not interacting with the coherent pulse following the emission of a photon into the waveguide.    
\begin{figure}
\centering
\includegraphics[scale = 1.0]{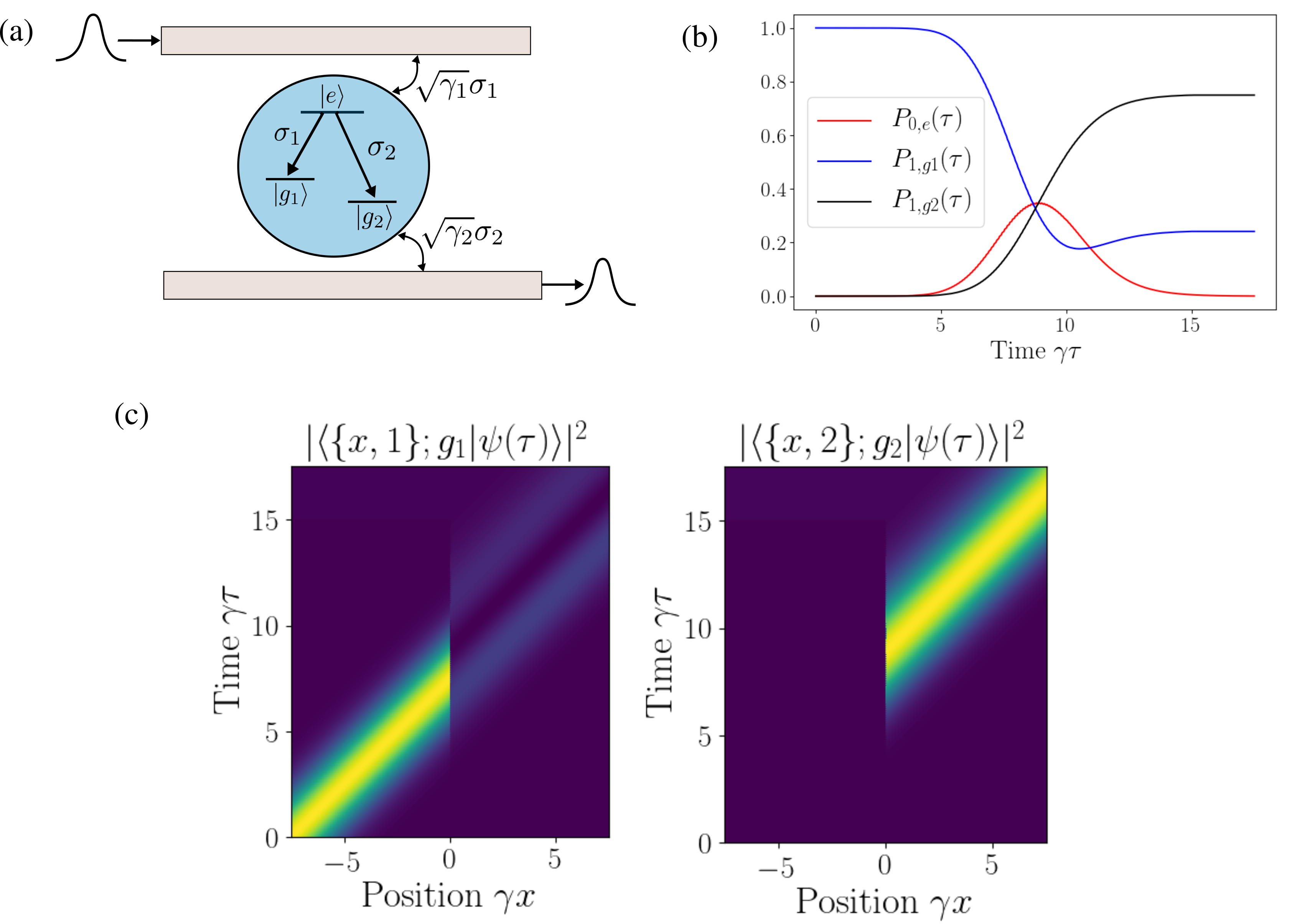}
\caption{Scattering of a single-photon wave-packet from a three-level lambda system. (a) Schematic of the system being analyzed. (b) Time dependence of the probabilities $P_e, P_{1, g_1}$ and $P_{2, g_2}$ for a three-level lambda system excited by an incident single-photon state. (c) Space-time dependence of the single-photon component of the system state in both the input and output waveguides. $\gamma_1 = \gamma_2 = \gamma/2$, $\Delta x = 2/\gamma$ and $\delta_e = \delta_{12} = 0$ is assumed in all the simulations.}
\label{fig:lambda_scattering}
\end{figure}

As our final example, we analyze photon subtraction using a lambda system --- the system under consideration is shown in Fig.~\ref{fig:lambda_scattering}(a). A lambda system is coupled to an input waveguide through the operator $\sigma_1$ and to an output waveguide through $\sigma_2$ to an output waveguide, with lambda system initially being in the state $\ket{g_1}$. A single photon is incident from the input waveguide which drives the system from $\ket{g_1}$ to $\ket{e}$ to $\ket{g_2}$, with the photon finally being emitted into the output waveguide. For a lambda system with $\delta_{12} = 0$ and the incoming photon being resonant with the ground state to excited state transition, the incoming photon is completely transfered from the input waveguide to the output waveguide. Subsequently, a photon incident from the input waveguide does not interact with the lambda system, since the lambda system is now in the state $\ket{g_2}$ which cannot be driven with an excitation from the input waveguide. This allows the lambda system to be used as a photon subtracter \cite{rosenblum2016extraction} --- for a stream of spatially separated single photons incident on the lambda system, the lambda system would remove the first photon from the stream, and transmit the rest. To numerically reproduce this effect, we consider the system to be in the following initial state:
\begin{align}
\ket{\psi(0)} = \int \psi_\text{in}(x_1)  a_{x_1,1}^\dagger \ket{\text{vac}; g_1} \textrm{d}x_1, \quad \psi_\text{in}(x_1) = \frac{1}{(\pi \Delta x^2)^{1/4}}\exp\bigg(-\frac{(x_1+L)^2}{2\Delta x^2} \bigg)
\end{align}
where $\Delta x$ is its spatial extent of the incoming photon wave-packet and $L$ is the distance of the center of the wave-packet from the lambda system at $t = 0$. The state of the system at $t = \tau$ can then be expressed in terms of the propagator matrix elements:
\begin{align}
\ket{\psi(\tau)} &= \int \bra{\text{vac}; e} U(\tau,0) \ket{\{x_1', 1\};g_1}\psi_\text{in}(x_1') \textrm{d}x_1' \nonumber \\ &\quad+  \int \int\bra{\{x_1,1\}; g_1} U(\tau,0) \ket{\{x_1', 1\};g_1}\psi_\text{in}(x_1') a_{x,1}^\dagger \ket{\text{vac}; g_1}\textrm{d}x_1' \textrm{d}x_1 \nonumber \\ 
&\quad+\int \int\bra{\{x_2,2\}; g_2} U(\tau,0) \ket{\{x_1', 1\};g_1}\psi_\text{in}(x_1') a_{x,2}^\dagger \ket{\text{vac}; g_2}\textrm{d}x_1' \textrm{d}x_2.
\end{align}
The probabilities of interest, denoted by $P_e, P_{1, g_1}$ and $P_{2, g_2}$, that we simulate for this system are defined by:
\begin{align}
P_e(\tau) = |\braket{\text{vac}; e | \psi(\tau)}|^2,\quad P_{1,g_1} = \int |\braket{\{x_1,1\}; g_1 | \psi(\tau)}|^2 \textrm{d}x_1 \quad \text{and} \quad P_{2,g_2} = \int |\braket{\{x_2,1\}; g_2 | \psi(\tau)}|^2 \textrm{d}x_2.
\end{align}

Figure~\ref{fig:lambda_scattering}(b) shows the time evolution of these probabilities --- clearly, the lamba-system transitions from initially being in $\ket{g_1}$ to the excited state $\ket{e}$ and then to $\ket{g_2}$ with photon transfering from the input to the output waveguide. However, we note that the photon wave-packet is not completely transmitted into the second waveguide, which is a consequence of wave-packet having a finite spread in frequency, and therefore not being completely resonant with the lambda system. Figure~\ref{fig:lambda_scattering}(c) shows the space-time dependence of the single-photon wave packet in the input and output waveguides --- clearly, the initial wave-packet propagates along the input waveguide till it reaches the lambda system at $x = 0$ and then transmits to the second waveguide.

\section{Conclusion and outlook}

\noindent We have completely described the unitary propagator for a composite quantum network comprising waveguides and a low-dimensional system. Our method is strongly connected to the linearity of the waveguide dispersions and the Markovian coupling approximation. This linearity allows us to always express the initial (time $\tau_-$) and final (time $\tau_+$) waveguide states in terms of Heisenberg field operators. As a result, we are able to use the input-output boundary conditions to arrive at an expression between the states in terms of system Heisenberg operators only. The culminating expectations are with respect to vacuum states, otherwise known as a Green's functions. These Green's functions can be expressed entirely in the Hilbert space of system operators, which we showed by coarse-graining the waveguides' spatial dimensions. Hence, our expression proves tractable for many systems of interest and provides insight into the connections between input-output theory, scattering matrices, and propagators for Markovian open-quantum systems.

\section{Acknowledgements}

We gratefully acknowledge financial support from the Air Force Office of Scientific Research (AFOSR) MURI Center for Quantum Metaphotonics and Metamaterials. RT acknowledges support from Kailath Stanford Graduate Fellowship. KAF acknowledges support from the Lu Stanford Graduate Fellowship and the National Defense Science and Engineering Graduate Fellowship.


\bibliography{bibliography}

\end{document}